\documentclass[aps,prb,twocolumn,showpacs,groupedaddress]{revtex4}
\usepackage{amsbsy,amssymb,amsmath,bm}
\usepackage{graphicx}

\input{epsf}

\begin{document}
\title {Generation of non-classical photon states in superconducting quantum metamaterials}
\author{S. I. Mukhin$^{1}$ and M. V. Fistul$^{1,2}$}

\affiliation {$^{1}$ Theoretical Physics and Quantum Technologies Department, Moscow Institute for Steel and Alloys, Leninski avenue 4, 119049 Moscow, Russia\\
$^{2}$ Theoretische Physik III, Ruhr-Universit\"at
Bochum, D-44801 Bochum, Germany }

\date{\today}
\begin{abstract}
We report a theoretical study of diverse non-classical photon states that can be realized in superconducting quantum metamaterials.
As a particular example of superconducting quantum metamaterials an array of SQUIDs incorporated in a low-dissipative transmission line (resonant cavity) will be studied. This system will be modeled as a set of two-levels systems (qubits) strongly interacting with resonant cavity photons.
We predict and analyze \emph{a second-order phase transition } between an incoherent (the high-temperature phase) and  coherent (the low-temperatures phase) states of photons.  In equilibrium state the partition function $Z$ of the electromagnetic field (EF) in the cavity is  determined by  the effective action $S_{eff}\{P(\tau)\}$ that, in turn, depends on imaginary-time dependent momentum of photon field $P(\tau)$. We show that the order parameter of this phase transition is the $P_{0}(\tau)$ minimizing the effective action of a whole system. In the incoherent state the order parameter $P_{0}(\tau)=0$ but at low temperatures we obtain various coherent states characterized by non-zero values of $P_{0}(\tau)$.
This phase transition in many aspects resembles the Peierls metal-insulator  and the metal-superconductor phase transitions.
The critical temperature of such phase transition $T^\star$ is determined by the energy splitting of two-level systems $\Delta$, a number of SQUIDs in the array $N$, and the strength of the interaction $\eta$ between SQUIDs and photons in cavity.
\\
\\
(submitted to Superconductor Science and Technology)
\end{abstract}


\maketitle

\section{Introduction}
A great attention is devoted to a theoretical and experimental study of novel quantum metamaterials \cite{QM1,QM2,QM3}. These systems consists of  a large amount of solid-states elements (qubits), i.e. two-levels systems showing  diverse coherent quantum phenomena, e.g. quantum beating, microwave induced Rabi oscillations, Ramsey fringes etc. \cite{qubits}. To obtain such coherent quantum-mechanical  behaviour in single qubits, the dissipation and decoherence have to be small enough \cite{qubits}. Moreover, in order to observe novel \emph{collective coherent quantum effects}  in metamaterials a strong long-range interaction between single elements has to be provided by surrounding media.

Various superconducting systems, e.g. arrays of Josephson junctions, RF SQUIDs, many-junctions superconducting quantum interferometers, just to name a few,  incorporating in low-dissipative (superconducting) transmission line are extremely suitable in order to realize such quantum metamaterials. Indeed, dc-biased Josephson junctions \cite{Martinis,Ustinov} or diverse SQUIDs \cite{Mooij,IBM} subject to an externally applied magnetic field have been established recently as qubits in which various coherent quantum effects have been observed. A strong long-range interaction between well-separated qubits is provided by a transmission line through emission (absorption) of virtual photons. This type of interaction was proposed in Refs. \cite{Wallr,Zagoskin,FU1,FU2,Fglstate} and realized in experiments with single qubits incorporated in a resonator \cite{Wallr2,Mooij2}. It has been shown that a strong interaction between well-separated qubits  results in an enhancement of quamtum-mechanical tunneling \cite{Fglstate,UstMuell,Nori} and suppression of decoherence induced by a spread of parameters of qubits \cite{Awsch}. The measurements of frequency dependent transmission (reflection) coefficient of electromagnetic field (EF) propagating through the transmission line provides a convenient method to observe coherent quantum phenomena in such metamaterials \cite{Abdumalikov,UstIl}.

On other hand a strong interaction of qubits with EF can result in different states of photons in the cavity. Indeed, in the absence of interaction with qubits the unique photon state of the cavity is \emph{an incoherent, chaotic state} of photons characterized by a well-known Planck distribution, i.e. $<\hat E >=0$, $<(\hat E)^2~\propto~[e^{-\frac{\hbar \omega_0}{k_BT}}-1]^{-1}>$, where $E$ is the electric field of radiation, $<...>$ is the quantum-mechanical average, and $\omega_0$ is the frequency of the cavity mode. A strong interaction of EF with qubits leads to the effective enhancement of energy levels difference of qubits that, in turn, changes EF in the cavity. Thus, one can expect that in the resonant cavity strongly interacting with an array of qubits the different states of photons can be observed. In this paper we  analyze possible quantum-mechanical states of photons emerging in the resonant cavity strongly interacting with an array of qubits. Notice here, that a similar analysis of the photon states of the cavity interacting with an unbiased array of Josephson junctions has been done in \cite{Stroud} in order to explain a strong radiation from a 2d-array of Josephson junctions observed in Ref. \cite{BarbaraSchitov}.

The paper is organized as follows: in Section II we present a model of an array of SQUIDs incorporating in a low-dissipative transmission line, and elaborate the classical description of this system, i.e. the dynamic equations of motion and the Lagrange function, in Section III  a complete quantum-mechanical description of this model is provided. In section IV using a great similarity with well-known phase transitions, e.g.  the metal-ferromagnet \cite{FMT}, metal-superconductor \cite{BCS} and the Peierls metal-insulator \cite{PT1} transitions, we predict and analyze a second-order phase transition between the incoherent, chaotic state (the high-temperature phase) of photons and diverse \emph{coherent non-classical} photon states (the low-temperature phase). The Section V provides discussion and conclusion.

\section{Model and classical description of superconducting metamaterials}
As a particular example we consider here a system of RF SQUIDs incorporated in a low-dissipative transmission line. An each RF SQUID is characterized by a Josephson phase $\varphi_i=2\pi \Phi_i/\phi_0$, where $\Phi_i$ is the total flux in the superconducting loop of a SQUID, and $\phi_0$ is the flux quantum. An application of dc-magnetic field characterized by $\Phi_{ext}$ allows one to tune the potential relief of a Josephson phase $\varphi_i$ from a single well up to a double-well potential. The set of RF SQUIDs is incorporated in a linear transmission line. The transmission line is characterized by two parameters $L_0$ and $C_0$, the inductance and capacitance per unit length, accordingly. We also introduce the voltage $V(x)$ and current $I(x)$ distributions, where $x$ is the coordinate along a transmission line. The inductive coupling, $M=\eta L_0$, provides an interaction between RF SQUIDs and transmission line. The schematic of a system is presented in Fig. 1.

\begin{figure}[tbp]
\includegraphics[width=1.5in,angle=-90]{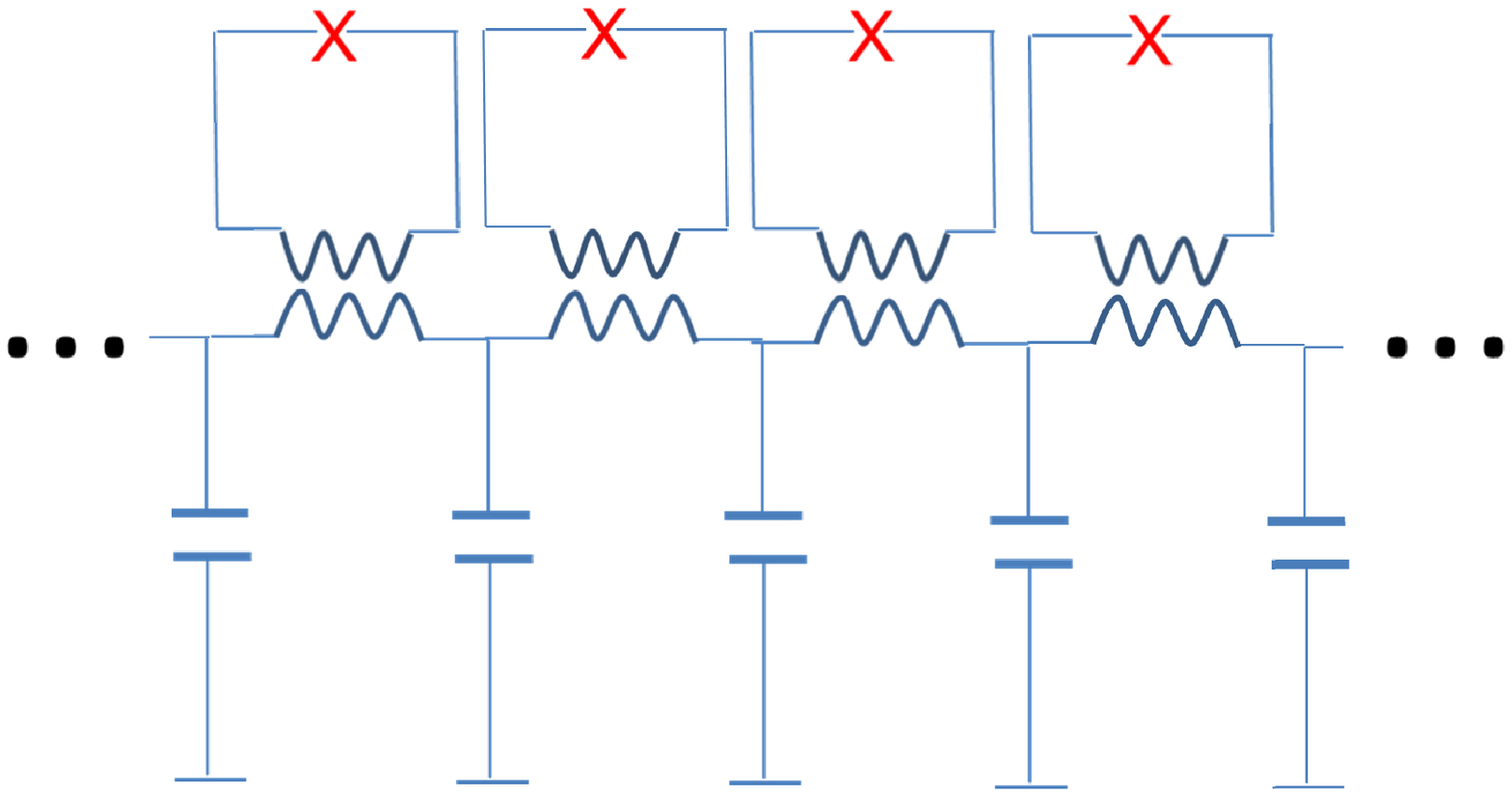}
\caption{The schematic of an array of RF SQUIDs incorporated in a transmission line.
} \label{schematic}
\end{figure}

\subsection{Classical Equations of Motion: linear transmission line}

We start with the classical dynamic equations for a linear transmission line. It is
  \begin{equation} \label{Voltage-eq}
\frac{\partial V(x,t)}{\partial x}=L_0\frac{\partial I(x,t)}{\partial t}
\end{equation}
and
 \begin{equation} \label{Current-eq}
\frac{\partial I(x,t)}{\partial x}=C_0\frac{\partial V(x,t)}{\partial t}
\end{equation}
These two equations are rewritten as
  \begin{equation} \label{EW-eq}
\frac{\partial^2 I(x,t)}{\partial x^2}=\frac{1}{c^2}\frac{\partial^2 I(x,t)}{\partial t^2}~,
\end{equation}
where $L_0C_0=1/c^2$. The electromagnetic standing waves can occur in this 1-d cavity resonator. The wave vectors are determined by standard boundary conditions: $k_n=\pi n/L$, where $L$ is the size of a transmission line, $n=1,2....$. We will consider a transmission line with an extremely high quality factor, which was routinely obtained in superconducting transmission lines, and therefore, the only one wave vector will be important in the dynamics of coupled RF SQUIDs and EWs of the cavity.

\subsection{Classical Equations of Motion: an individual SQUID}
The classical dynamic equations for an individual SQUID are written as
   \begin{equation} \label{Eq-motion-SQUID}
I_{RF}/I_c=\sin(\varphi)+\frac{\alpha}{\omega_p} \frac{d\varphi}{dt}+\frac{1}{\omega_p^2}\frac{d^2\varphi}{dt^2}~~,
\end{equation}
where $I_c$ is the critical current of a Josephson junction, $1/\alpha^2$ is the McCumber parameter characterizing the dissipation of SQUID, $\omega_p$ is the plasma frequency of a Josephson junction. On other hand the Josephson phase in a SQUID loop is satisfied to the following equation as
\begin{equation} \label{Eq-motion-SQUID-2}
\varphi=\varphi_{ext}-\beta_LI_{RF}/I_c~,
\end{equation}
where $\beta_L$ is the inductive (dimensionless) parameter of the SQUID, $\varphi_{ext}$ corresponds to the sum of the externally applied dc-magnetic field and ac-magnetic field induced by a current flowing along the transmission line. Thus, $\varphi^{dc}_{ext}$ allows one to tune the potential relief of the Josephson phase, and $\varphi^{ac}_{ext}~\propto~I$ provides a coupling between the RF SQUID and a transmission line.

\subsection{Classical Equation of Motion: coupled transmission line and SQUIDS}
Inductive coupling between RF SQUIDs and transmission line results into a particular change of classical equations of motion: first, the Eq. (\ref{EW-eq}) changes to
\begin{equation} \label{EW-eq-2}
\frac{\partial^2 I(x,t)}{\partial x^2}=\frac{1}{c^2}\frac{\partial^2 I(x,t)}{\partial t^2}+\sum_i \frac{\eta}{c^2}\frac{\partial^2 I^{(i)}_{RF}(t)}{\partial t^2},
\end{equation}
 where $\eta$ is the parameter characterizing a mutual inductance of RF SQUIDs and the transmission line; secondly, $\varphi^{ac}_{ext}=(\eta L_0/\phi_0) I(x_i,t)$.

 \subsection{Lagrangian of a Superconducting Metamaterial}

 The classical equations of motion can be derived also from the Lagrangian of a whole system
 $$
 L=\frac{L_0}{2}(\dot Q(x,t))^2-\frac{1}{2C_0}(\frac{\partial Q(x,t)}{\partial x})^2+
 $$
 $$
 +E_J\sum_i \frac{1}{2\omega_p^2}[\dot \varphi_i]^2-\frac{1}{2\beta_L}\left[\varphi_i- \varphi^{dc}_{ext}+(\eta L_0/\phi_0) \dot Q(x,t)\right]^2-
 $$
 \begin{equation} \label{Lagrangian}
-(1-\cos \varphi_i),
\end{equation}
where $Q(x,t)$ is the charge variable characterizing a transmission line, and $E_J=\hbar I_c/(2e)$ is the energy of Josephson junction.
To simplify this expression we consider the standing EWs with a single wave vector $k_n$ only. In this case the charge variable $Q(x,t)$ has a form:
$Q(x,t)=Q(t)\cos(k_n x)$. Substituting this expression in Eq. (\ref{Lagrangian}) we obtain
$$
L=L_{ph}+L_{SQUID}+L_{int}~,
$$
where
\begin{equation} \label{Lagrangian-photon}
 L_{ph}=m\left[\frac{1}{2}(\dot Q(t))^2-\frac{c^2 k_n^2}{2}(Q(t))^2\right]~, ~m=L_0/2,
\end{equation}
and
\begin{equation} \label{Lagrangian-RFSQUIDS}
L_{SQUID}=E_J\sum_i \frac{1}{2\omega_p^2}[\dot \varphi_i]^2-\frac{1}{2\beta_L}[\varphi_i- \varphi^{dc}_{ext}]^2-(1-\cos \varphi_i),
\end{equation}
and
\begin{equation} \label{Lagrangian-interaction}
L_{int}=-\sum_i (\eta L_0/\Phi_0)\frac{E_J}{\beta_L}\varphi_i \cos(k_n x_i)\dot Q(t),
\end{equation}
where $x_i$ are the coordinates of RF SQUIDs along a transmission line.

\section{Quantum Description of Superconducting Quantum Metamaterials}

\subsection{Photon Hamiltonian}
Introducing the "momentum " of a photon field $P(t)$ as $P=(L_0/2)\dot Q(t)$ we obtain the Hamiltonian of a photon field in the following form:
\begin{equation} \label{Hamiltonian-Photon}
H_{ph}=\frac{P^2}{L_0}+\frac{L_0 c^2 k_n^2 }{4}[Q(t)]^2~.
\end{equation}
Next we introduce the operators of  boson field $\hat b$ and $\hat b^+$ as
$$
\hat Q(t)=\sqrt{\frac{\hbar}{ck_nL_0}}(\hat b+\hat b^+)
$$
and
$$
\hat P(t)=-i\sqrt{\frac{\hbar ck_nL_0}{4}}(\hat b-\hat b^+)~.
$$
Using these new variables the photon Hamiltonian is written as
\begin{equation} \label{Hamiltonian-Photon-2}
H_{ph}=\hbar  \omega_0(\hat b^+ \hat b+1/2)~~, \omega_0=ck_n.
\end{equation}

\subsection{RF SQUID Hamiltonian}
We consider the macroscopic quantum dynamics of RF SQUID when the potential energy has a form of double well potential. In this case the Hamiltonian of a system of isolated RF SQUIDs is written as
\begin{equation} \label{Hamiltonian-RFSQUID}
H_{SQUID}=\sum_i \frac{1}{2}[\Delta_i \hat \sigma_x+\epsilon_i \hat \sigma_z]~,
\end{equation}
where $\sigma_z$ and $\sigma_x$ are standard Pauli matrices, $\Delta_i$ and $\epsilon_i$ are the matrix element of tunneling and the energy difference between two potential wells, accordingly. Notice here, that such a Hamiltonian can be used also for more complex qubits, e.g. phase qubits, flux qubits etc, where parameters $\Delta_i$ and $\epsilon_i$ are determined by the physical properties of corresponding qubits. For qubits based on RF SQUIDs we obtain \begin{equation} \label{RFQubit-delta}
\Delta ~\propto~\hbar \omega_p (1-1/\beta_L)^{1/2} e^{-\frac{4\sqrt{2}E_J}{\hbar \omega_p}(1-1/\beta_L)^{3/2}},
\end{equation}
and
\begin{equation} \label{RFQubit-epsilon}
\epsilon~\propto~\pi \sqrt{6}(1-1/\beta_L)^{1/2} (\phi_{ext}/\phi_0-1).
\end{equation}
Moreover, the parameters $\Delta_i$ and $\epsilon_i$ can fluctuate from one qubit to other one.

\subsection{Interaction Hamiltonian}
The equilibrium dynamics of a Josephson phase in imaginary time representation can be presented as rare jumps (the instanton type of solution) between two equilibrium positions \cite{Chakr}. Using this property we obtain the interaction Hamiltonian as follows:
 \begin{equation} \label{Hamiltonian-Interaction}
H_{int}=i\sum_i \xi_i\hat \sigma_z (\hat b-\hat b^+)~,~\xi_i=\frac{E_J\eta (\delta \varphi) \sqrt{\hbar ck_nL_0}}{\beta_L\phi_0}
\end{equation}
where $\delta \varphi$ is the phase difference between two equilibrium positions of a Josephson phase.

\subsection{Effective action}
In order to study the various photon states arising in superconducting quantum metamaterials we write the partition function $Z$ in the form of functional integral as
 \begin{equation} \label{Partitionfunction}
Z=\int DQD\{\varphi_i\}\exp({-S\{Q,\varphi_i \}}),
\end{equation}
 where $S$ is the action of EF interacting with an array of two-levels systems. Integrating the Eq. (\ref{Partitionfunction}) over $\{\varphi_i\}$ we obtain the effective action $S_{eff}$ in the following form \cite{Mukhin}:
 $$
 S_{eff}[Q(\tau)]=\frac{1}{\hbar}\int_0^{\hbar/(k_BT)}d\tau \frac{m}{2}\left[{\dot Q}^2+\omega_0^2 Q^2\right]-
 $$
  \begin{equation} \label{effectiveaction}
-k_BT\sum_i \ln \left[\cosh \frac{\alpha_i\{Q\}}{k_BT}\right],
\end{equation}
where $\alpha_i \{Q(\tau)\}$ are the positive Floquet eigenvalues of arrays of two-levels systems in the presence of \emph{periodic in imaginary time potential } $Q(\tau)$. Notice here, that in the absence of interaction  with the field $Q(\tau)$, i.e. as $\xi_i=0$, the Floquet eigenvalues are $\alpha_i(0)=\sqrt{\Delta_i^2+\epsilon_i^2}$, and the minimum of $S_{eff}[Q(\tau)]$ occurs as $Q=0$.

\section{Phase transitions in states of photons}

Since the interaction term in the Lagrange function $L_{int}$ depends on $\tau$-derivative of $Q$, i.e. $\dot Q$, we introduce a new variable of the momentum of a photon field $P(\tau)=m \dot Q$ and rewrite the effective action $S_{eff}$ as
$$
 S_{eff}[P(\tau)]=\frac{1}{\hbar}\int_0^{\hbar/(k_BT)}d\tau \frac{1}{2m}\left[P^2+\frac{1}{\omega_0^2} \dot P^2\right]-
 $$
  \begin{equation} \label{effectiveaction-P}
-k_BT\sum_i \ln \left[\cosh \frac{\alpha_i\{P\}}{k_BT}\right]~~.
\end{equation}
Moreover, the Floquet eigenvalues $\alpha_i(P)$ are determined from the following equation:
\begin{equation} \label{Equation-Floquet}
(\partial_\tau+\hat H^{i}_{P})\psi_i=0
\end{equation}
and the Hamiltonian $H^{i}_{P}$ is
\begin{equation} \label{Hamiltonian-Floquet}
\hat H^{i}_{P}=\left(
    \begin{array}{cc}
      \epsilon_i+\tilde \eta_i P & \Delta_i\\
      \Delta_i & -(\epsilon_i+\tilde \eta_iP)
    \end{array}
  \right)~,
\end{equation}
where the parameter determining the interaction between cavity modes and two-levels systems $\tilde \eta_i = \frac{\eta (\delta \varphi_i)E_J}{\Phi_0\beta_L} \cos(k_n x_i)$.

Next, we obtain the periodic in imaginary time representation function $P_{0}(\tau)$ minimizing the effective action, as a solution of the following equation:
\begin{equation} \label{Min-effectivaction}
\frac{P_0}{m}+\frac{1}{m\omega_0^2}\ddot P_0=\sum_i \frac{\partial \alpha_i}{\partial P} tanh \left(\frac{\alpha_i\{P_0\}}{k_BT}\right)~.
\end{equation}

\subsection{Classical Second Order Phase Transition}

First we consider a particular case as the momentum of a photon field $P_0$ \emph{does not} depend on $\tau$. The Floquet eigenvalues are determined as $\alpha_i(P)=\sqrt{(\epsilon_i+\tilde \eta_i P_0)^2+\Delta_i^2}$ and the equation for $P_0$ reads as
\begin{equation} \label{Eq-P0}
\frac{P_0}{m}=\sum_i \tilde \eta_i \frac{(\epsilon_i+\tilde \eta_i P_0)}{\sqrt{(\epsilon_i+\tilde \eta_i P_0)^2+\Delta_i^2}} tanh \left[\frac{\sqrt{(\epsilon_i+\tilde \eta_i P_0)^2+\Delta_i^2}}{k_BT}\right]
\end{equation}
In a simplest case as $\epsilon_i=0$ and $\tilde \eta_i=\tilde \eta $ we obtain self-consistent equation
\begin{equation} \label{SelfconsEq-P0}
P_0\left[\frac{1}{m\tilde \eta^2}-\sum_i \frac{1}{\sqrt{\tilde \eta^2 P_0^2+\Delta_i^2}} tanh \left(\frac{\sqrt{\tilde \eta^2 P_0^2+\Delta_i^2}}{k_BT}\right)\right]=0
\end{equation}
At high temperatures, $T>T^{\star}$, this equation has a single solution $P_0=0$. Such a high-temperature phase corresponds to the incoherent, chaotic state of a photon field. However, at low temperatures, $T<T^{\star}$, the Eq. (\ref{SelfconsEq-P0}) has two non-zero solutions of $\pm P_0$, and \emph{the coherent state} of a photon field can be realized. Therefore, we obtain \emph{a second-order type of phase transition} in the states of photon field interacting with a set of two-levels systems. The critical temperature $T^{\star}$ depends on the distribution of energy differences of two-levels system $\Delta_i$. If such distribution is a narrow one, i.e. $\Delta_i \simeq \Delta$, we obtain the value of $T^\star$ as
\begin{equation} \label{Temp-PhaseTransition}
T_{n}^\star =\frac{m\tilde \eta^2 N}{k_B}~,
\end{equation}
where $N$ is a total number of two-levels systems. Such a phase transition occurs only if $\Delta_0 < k_B T_n^\star$. At low temperatures  $P_0$ reaches the maximum value of $P_0=\pm m\tilde \eta N$. As one can see the maximum value of $P_0$ is proportional to $N$, and this dependence also indicates the presence of the coherent state. The dependence of $P_0(T)$ for this case is shown in Fig.2 (red solid line). Notice here, that this case resembles a well-known metal-ferromagnet phase transition \cite{FMT}.

In the opposite case of a wide distribution of $\Delta_i$, e.g. from zero to upper cut-off $\Delta_0$,  the critical temperature is determined as
\begin{equation} \label{Temp-PhaseTransition-wide}
T_{w}^\star =\Delta_0 e^{-\frac{\Delta_0}{m\tilde \eta^2 N}}~.
\end{equation}
This phase transition occurs if $\Delta_0> m\tilde \eta^2 N$. At low temperatures the $P_0$ reaches the maximum value of $|P_0|~\simeq~k_B T_w^\star/\tilde \eta$.  This phase transition resembles the superconductor-normal metal \cite{BCS} and/or Peierls metal-insulator \cite{PT1} transitions.

\begin{figure}[tbp]
\includegraphics[width=2.8in,angle=0]{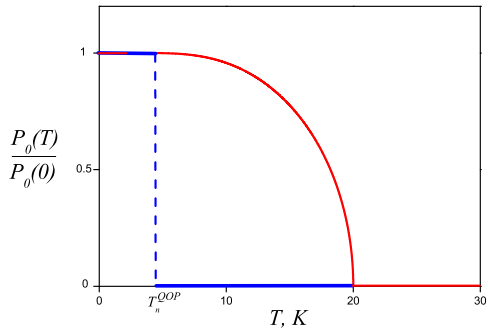}
\caption{The classical and quantum phase transitions in the state of photons. The temperature dependencies of the momentum of photon field $P_0(T)$ are shown: $\tau$-independent $P_0$ (red line) and $P_0(\tau)~\propto~P_0(T)\sin(P_0(T) \tau)$ (blue line). The case of a narrow distribution of $\Delta_i \simeq ~\Delta_0$ is shown. The parameters $T_n^\star=20 K$, and $\Delta_0=4 K$ were used.
} \label{schematic}
\end{figure}

\subsection{Quantum Phase Transitions}
Here we consider a phase transition into a peculiar "quantum ordered" state representing a quantum interference between the two semi-classical states  $\pm P_0$ of the photon field, which are inversion symmetry related solutions of the self-consistency equation Eq.(\ref{SelfconsEq-P0}). Each of the $\pm P_0$ states separately describes a particular coherent state of the photon field in our model. The quantum ordered state was predicted in \cite{Mukhin} for a system of electron-hole pairs  coupled to a semi-classical spin-density wave fluctuations. In our present model it is described by the amplitude of the semi-classical photon field $P_0(\tau)$, or the "quantum order parameter", which is the periodic in imaginary time solution of Eq. (\ref{Min-effectivaction}). Again we consider a simplest case as $\epsilon_i=0$ and $\tilde \eta_i=\tilde \eta $ and  apply analytic solution found in \cite{Mukhin}:
\begin{equation} \label{Selfcons-tau-P0}
\tilde \eta P_0(\tau)= 4nKTk_1 sn\left( {4nKT\tau ;k_1 } \right),\,K = K(k_1 )
\end{equation}
where $sn(\tau,k_1)$ is the Jacobi snoidal elliptic function, periodic in $\tau$ with period $1/(nT)$, $n=1,2,...$, $K(k_1 )$ is elliptic integral of the first kind, positive integer $n$ and  real parameter $0< k_1 <1$ are found by minimizing the Euclidian action $S_{eff}$ given in Eq. (\ref{effectiveaction-P}). Here we merely describe a single-harmonic limit $k_1\rightarrow 0$ of solution Eq.(\ref{Selfcons-tau-P0}). In this limit expression for $P_0(\tau)$ in Eq.(\ref{Selfcons-tau-P0}) turns into:
\begin{equation} \label{Sin-tau-P0}
\tilde \eta P_0(\tau)\approx 2\pi nTk_1 sin\left( 2\pi nT\tau \right).
\end{equation}
\noindent Simultaneously, it was shown in \cite{Mukhin1}, that solution in the form of Eq.(\ref{Selfcons-tau-P0}) leads to the following spectrum of the Floquet eigenvalues $\alpha_i(P)$ found from Eq.(\ref{Equation-Floquet}) :
\begin{equation} \label{specpi}
{\alpha}_i=2\tilde\Delta_i\left(\dfrac{1-{{k}}^2+\tilde\Delta_i^2}{1+\tilde\Delta_i^2}\right)^{1/2}
n\Pi\left(\dfrac{{{k}}^2}{1+\tilde\Delta_i^2},{{k}}\right)
\end{equation}
\noindent Here $\Pi(m,k)$ is elliptic integral of the third kind, and besides:
\begin{equation}
\tilde\Delta_i\equiv \dfrac{\Delta_i}{2TnK(k)},\;{k}=2\sqrt{k_1}/(1+k_1); \,k'^2={1-k^2}
\label{renorm}
\end{equation}
The latter relation between parameters $k_1,k$ is known as Landen transformation \cite{witt}. The Jacobi's function $M(\tau)=M(\tau+1/nT)$ from Eq. (\ref{Selfcons-tau-P0}) turns the generic {\it{self-cosistency equation}} (\ref{Min-effectivaction}) into algebraic equation for parameters $k,n$ \cite{Mukhin}:
\begin{equation} \label{QOP-self}
\sum_{i} \left[\tanh\dfrac{{\alpha}_i}{k_BT}\right]\dfrac{\tilde\Delta_i}{\{(\tilde\Delta_i^{2}+ 1)(\tilde\Delta_i^{2}+k'^2)\}^{1/2}}=\dfrac{1}{m\tilde \eta^2}
\end{equation}
It is not hard to see, that in the limit $k'\rightarrow 0$ Eq.(\ref{QOP-self}) transforms into classical mean-field self-consistency Eq. (\ref{SelfconsEq-P0}). While in the limit $k'\rightarrow 1$ the self-consistency equation  Eq. (\ref{QOP-self}), as it follows from Eq.(\ref{specpi}) and Eq.(\ref{renorm}) turns into the following equation:
\begin{equation} \label{QOP-self-k0}
\sum_{i} \left[\tanh\dfrac{\Delta_i}{k_BT}\right]\dfrac{\Delta_i}{\Delta_i^{2}+(\pi nT)^2}=\dfrac{1}{m\tilde \eta^2}
\end{equation}
Now, after a comparison with the Eq.(\ref{SelfconsEq-P0}) for the classical photon condensate $P_0$ it is possible to conclude, that in the case of narrow distribution of energy differences of the two-levels systems $\Delta_i\approx\Delta_0$ and strong coupling constant to the electromagnetic field: $\Delta_0< m\tilde \eta^2 N$, the quantum ordered phase (QOP) of the photon field occurs below the temperature:
\begin{equation} \label{QOP-PhaseTransition-n}
T_{n}^{QOP} \propto \left[\frac{\Delta_0^2 m\tilde \eta^2 N}{\pi^2}\right]^{1/3}\frac{1}{k_B}~,
\end{equation}
Since the amplitude of $P_0(\tau)$ is proportional to $4nKTk_1$ in accord with Eq. (\ref{Selfcons-tau-P0}), this result suggests that the number of photons condensed into quantum ordered phase is $\propto N^{1/3}$, where $N$ is the number of the two-lewel systems. Hence, this dependence indicates the presence of the coherent state also in the quantum ordered phase, but with less strong entangling than in the classical photon condensate described by Eq. (\ref{SelfconsEq-P0}).
In the opposite case of a wide distribution of $\Delta_i$ and weak coupling constant $\Delta_0> m\tilde \eta^2 N$ we find transition temperature similar to the classical mean-filed case Eq. (\ref{Temp-PhaseTransition-wide}):
\begin{equation} \label{QOP-PhaseTransition-wide}
T_{w}^{QOP} =\frac{\Delta_0}{\pi} e^{-\frac{\Delta_0}{m\tilde \eta^2 N}}~.
\end{equation}
In both cases, when temperature lowers well below $T_{n,w}^{QOP}$ the increase of the integer number $n\propto 1/T$ of the oscillations of the quantum order parameter $P(\tau)$ along the imaginary time-axis $\tau$ within the Euclidian space temporal interval $[0,1/T]$  keeps the QOP amplitude $4nKTk_1$ finite and non-vanishing up to $T=0K$ state acording to Eq.(\ref{Selfcons-tau-P0}). Thus, we can conclude that the quantum phase transition in the photon states is the first-order type of phase transition.

\section{Discussion and conclusions}
In previous Section we obtained classical and quantum second-order phase transitions in the states of photons that can be emerged in a resonant cavity strongly interacting with an array of two-levels systems. For the classical phase transition we obtain that at low temperatures, $T<T^\star$, the incoherent photon state (with $P_0=0$) becomes unstable. Two novel  photon states characterized by non-zero values of the momentum of photon field $\pm P_0(T)$ can appear at low temperatures. These photon states are the \emph{coherent photon states} well-known in the quantum optics \cite{Loudon}. The average number of photons in these states $\bar{n}=P_0^2/(2m \hbar \omega_0)$. The temperature dependence of $P_0(T)$ is determined by Eq. (\ref{SelfconsEq-P0}), and it is shown in the Fig. 2 (red solid line). The critical temperature $T^\star$ depends strongly on the distribution of energy differences $\Delta_i$, a number of qubits $N$, and the strength of interaction $\tilde \eta$. Thus, for reasonable values of parameters $T^\star$ varies from $100 mK$ up to $50 K$. The value of $T^\star$ is determined by Eqs. (\ref{Temp-PhaseTransition}) and (\ref{Temp-PhaseTransition-wide}).  This phase transition is similar to a well-known metal-ferromagnet transition, and the momentum of photon field $\pm P_0(T)$ corresponds to the "Weiss effective magnetic field" in a theory of ferromagnetic materials \cite{FMT}.

Notice here, that these two coherent photon states corresponding to the $+P_0(T)$ and $-P_0(T)$ values are the degenerate ones. In this case the quantum beating between these two states can be observed in the system. The quantum beating between two photon states results in a splitting of resonant frequencies of the cavity, i.e. $\omega_{res}=\omega_0 \pm \Delta \omega$. The splitting $\Delta \omega$ is obtained as follows:    the two stable photon states are separated by the effective potential barrier $\Delta U~\simeq~m N^2 \tilde \eta^2$ (for $T=0$), and therefore, $\Delta \omega~\simeq~\omega_0 \exp[-\Delta U/(\hbar \omega_0)]$. The effect of the splitting of resonant frequencies is a consequence of the degeneracy of the photon states. Such a degeneracy can be lifted by application of an external magnetic field allowing one to realize non-symmetric double-well potential for RF SQUIDs, i.e. as $\epsilon_i \neq 0$. In this case a single coherent state of photons with $P_0~\propto~\sum_i \epsilon_i$ emerges in the cavity.

We obtained also that different metastable states of photons can be obtained in this system. In these states there is no net classical photon condensate, but there exists macroscopic quantum condensate (with amplitude of the photon momentum $P\propto N^{1/3}$, where $N$ is the number of two-level systems), that has zero mean value of the electromagnetic field. These states appear as a result of a first-order phase transition.

In conclusion we have shown that superconducting quantum metamaterials can support the diverse non-classical photon states. As a particular example of such metamaterial we considered an array of RF SQUIDs incorporated in a low-dissipative resonant cavity. We mapped this system to a set of two-levels systems (qubits) strongly interacting with photons of the cavity. By making use of a complete quantum-mechanical description of such a system we obtained that at high temperatures, $T>T^\star$ the incoherent chaotic state of photons is a stable one. At low temperatures $T<T^\star$ a large amount of different photon states emerges in the cavity. These photon states appear as a result of  specific classical (the second-order type) or quantum (the first-order type) of phase transition.
The physical origin of such phase transitions is a following: a strong interaction of EF with two-levels systems leads to the effective enhancement of energy levels difference of qubits that, in turn, changes  the EF in the cavity. The order parameter of phase transitions is the $\tau$-dependent momentum of photon field $P_0(\tau)$. In the case of the classical phase transition as $P_0(\tau)=const$ the \emph{coherent photon states } and  the quantum superposition of two coherent photon states can be obtained in the cavity. In the case of quantum phase transitions the different metastable photon states characterized by complex dependence of $P(\tau)$ (see Eq. (\ref{Selfcons-tau-P0})) also can be realized.

We acknowledge partial support of this work by the Russian Ministry of science and education grants No. 14A18.21.1936 and No. 11.G3431.0062.  S.I.M. also acknowledges partial support from the MISIS grant 3400022 and RFFI grant 12-02-01018.
{}

\end{document}